\documentclass{article}
\usepackage{sylvie}

\DeclareRobustCommand{\stir}{\genfrac\{\}{0pt}{}}

\title{State Complexity of Reversals of Deterministic Finite Automata with Output}
\author{Sylvie Davies}
\date{\small University of Waterloo\\Department of Pure Mathematics\\{\tt sldavies@uwaterloo.ca}}

\begin{document}
\maketitle
\begin{abstract}
We investigate the worst-case state complexity of reversals of deterministic finite automata with output (DFAOs). In these automata, each state is assigned some output value, rather than simply being labelled final or non-final. This directly generalizes the well-studied problem of determining the worst-case state complexity of reversals of ordinary deterministic finite automata. If a DFAO has $n$ states and $k$ possible output values, there is a known upper bound of $k^n$ for the state complexity of reversal. We show this bound can be reached with a ternary input alphabet. We conjecture it cannot be reached with a binary input alphabet except when $k = 2$, and give a lower bound for the case $3 \le k < n$. We prove that the state complexity of reversal depends solely on the transition monoid of the DFAO and the mapping that assigns output values to states. 
\end{abstract}
\section{Introduction}
Formal definitions are postponed until Section \ref{sec:def}.

The problem of determining the worst-case state complexity of the reversal operation on regular languages has been well-studied. Work on this problem dates back to the 1960s; see Jir\'askov\'a and \v{S}ebej~\cite{JiSe12} for a historical overview. It is known that if $L$ is recognized by an $n$-state deterministic finite automaton (DFA), then the (deterministic) state complexity of the reverse $L^R$ is at most $2^n$, and this bound can be reached over a binary alphabet; furthermore, it can be reached by DFAs which have only one final state.

In this paper, we study a generalization of this problem to \e{deterministic finite automata with output} (DFAOs). Rather than a set of final states, in a DFAO, each state is assigned an output value from a finite \e{output alphabet} $\Delta$. Rather than recognizing languages, DFAOs compute functions $f \co \Sig^* \ra \Delta$, where $\Sig$ is the \e{input alphabet}. The value $f(w)$ is defined to be the output value of the state reached by starting in the initial state and following the path corresponding to the input word $w$. Note that the case $|\Delta| = 2$ can be viewed as assigning a value of ``final'' or ``non-final'' to each state, so DFAOs directly generalize DFAs.

DFAOs are used in the study of \e{automatic sequences}~\cite{AlSh03}. If we treat the words $w \in \Sig^*$ as representations of natural numbers in some base, we can view the function $f \co \Sig^* \ra \Delta$ as a function $f \co \bN \ra \Delta$, that is, an infinite sequence of elements of $\Delta$. Sequences for which the corresponding function can be computed by a DFAO are called \e{automatic}. 

The \e{reverse} of the function $f \co \Sig^* \ra \Delta$ is the function $f^R \co \Sig^* \ra \Delta$ defined by $f^R(w) = f(w^R)$. The reversal operation on DFAOs can be thus be viewed as changing the direction in which the DFAO reads input: from left-to-right to right-to-left, or vice versa. 
Some functions are easier to compute with respect to one input-reading direction than the other. For example, consider the function $f \co \{0,1\}^* \ra \{0,1\}$, which takes in the binary representation of a natural number and outputs $1$ if the number can be written as $8n+5$, $n \ge 0$, and $0$ otherwise. Numbers of the form $8n+5$ have binary representations of the form $w101$, where $w \in \{0,1\}^*$. Hence if the input is read from left-to-right, the entire string must be read to determine whether it ends in $101$, but if the input is read from right-to-left, only three characters need to be checked. Likewise, some automatic sequences are easier to generate if we read the input numbers from least-significant digit to most-significant-digit, rather than the opposite way. 

We are concerned with the maximal blow-up in size (number of states) when the input reading direction of a DFAO is reversed. That is, given a function $f$ computed by an $n$-state DFAO, what is the worst-case state complexity of $f^R$? The standard construction for reversal of DFAOs~\cite[Theorem 4.3.3]{AlSh03} gives an upper bound of $|\Delta|^n$, where $\Delta$ is the output alphabet. However, it does not seem to be known whether this bound is reachable.

We prove that when the input alphabet has size three or greater, the upper bound $|\Delta|^n$ is indeed reachable. When the input alphabet is binary, the problem becomes much more complicated. 
We conjecture that if $|\Delta| \ge 3$, the upper bound $|\Delta|^n$ is not reachable over a binary alphabet, despite the fact that it is known to be reachable for $|\Delta| = 2$ (the ordinary DFA case). While we could not prove that the upper bound is unreachable in all cases, we have proved it is unreachable when $|\Delta| = n$  and $|\Delta| \ge 3$, and verified computationally that it is unreachable for $(|\Delta|,n) \in \{(3,4),(3,5),(3,6),(4,5)\}$.
We prove a lower bound for the case of a binary input alphabet and $3 \le |\Delta| < n$. We provide some preliminary computational evidence showing that this bound may be optimal for $n \ge 7$.

We also demonstrate that the state complexity of DFAO reversal is completely determined by the transition monoid of the DFAO and the map which assigns outputs to states. In particular, if function $f$ is computed by a minimal $n$-state DFAO with state set $Q$, transition monoid $M$, and output map $\tau \co Q \ra \Delta$, then the state complexity of $f^R$ is exactly $|\tau M|$, where $\tau M = \{ \tau \circ m : m \in M\}$ and $\circ$ denotes function composition. Since DFAs are special cases of DFAOs, this gives a surprising new characterization of the state complexity of DFA reversal in terms of the transition monoid and the characteristic function of the final state set.

\section{Preliminaries}
\label{sec:def}
We assume familiarity with basic concepts and results from the theory of formal languages and automata, particularly those related to regular languages and deterministic finite automata (DFAs). For example, see Hopcroft and Ullman~\cite{HoUl79}.

A \e{deterministic finite automaton with output} (DFAO) is a 6-tuple $\cD = (Q,\Sig,\cdot,q_0,\Delta,\tau)$, where:
\bi
\item
$Q$ is a finite set of \e{states} and $q_0 \in Q$ is the \e{initial state}.
\item
$\Sig$ is the \e{input alphabet} and $\Delta$ is the \e{output alphabet}; both are finite.
\item
$\cdot \co Q \times \Sig \ra Q$ is the \e{transition function}.
\item
$\tau \co Q \ra \Delta$ is the \e{output map}.
\ei
We use infix notation for the transition function: the image of the pair $(q,a)$ under the transition function is denoted $q \cdot a$.
We extend the transition function to words in $\Sig^*$ as follows: for $q \in Q$, we define $q \cdot \eps = q$, and for $w = ax$, $a \in \Sig$, $x \in \Sig^*$ we inductively define $q \cdot ax = (q \cdot a) \cdot x$.
If $p \cdot a = q$ for $p,q \in Q$ and $a \in \Sig$, we say there is a \e{transition} from $p$ to $q$ on $a$.
If $p \cdot w = q$ for $w \in \Sig^*$, we say there is a \e{path} from $p$ to $q$ on $w$.

While traditional DFAs recognize languages, DFAOs instead compute functions.
The \e{function computed by a DFAO} is the function $f \co \Sig^* \ra \Delta$ defined by $f(w) = \tau(q_0 \cdot w)$. That is, we determine $f(w)$ by starting in the initial state $q_0$, following the path corresponding to $w$ to reach some state $q$, then applying the output map $\tau$ to get the output value associated with $q$. A function that can be computed by a DFAO is called a \e{finite state function}.

A state $q \in Q$ is \e{reachable} if there is a path to it from the initial state $q_0$, i.e., there exists $w \in \Sig^*$ such that $q_0 \cdot w = q$.
The DFAO $\cD$ is called \e{trim} if all states are reachable.
Two states $p,q \in Q$ are \e{distinguishable} if there exists $w \in \Sig^*$ such that $\tau(p \cdot w) \ne \tau(q \cdot w)$. 
A DFAO is \e{minimal} if it has the least possible number of states among all DFAOs computing the same function. The following result is well-known for DFAs, and it can be shown to hold for DFAOs using essentially the same proof.
\bp
\label{prop:min}
A DFAO is minimal if and only if all states are reachable and every pair of distinct states is distinguishable.
\ep
For further reference on the DFAO model, see Allouche and Shallit~\cite{AlSh03}.

Let $Q$ be a finite set; we usually assume without loss of generality that $Q = \{1,2,\dotsc,n\}$. A \e{transformation} of $Q$ is a function $t \co Q \ra Q$. 
The \e{image} of a transformation $t \co Q \ra Q$ is the set $t(Q) = \{ t(q) : q \in Q \}$.
The \e{rank} of a transformation is the size of its image.
Transformations of $Q$ 
(or more generally, functions $f \co Q \ra X$ for some set $X$)
can be specified explicitly using \e{matrix notation}:
\[ t = \begin{pmatrix}
1&2&3&\dotsb&n\\\
t(1)&t(2)&t(3)&\dotsb&t(n)\end{pmatrix}.
\]
Transformations (or functions $f\co Q \ra X$) can be written concisely using \e{list notation}; for example, the list $[1,4,3,5,2,2,3]$ denotes the transformation
\[ \begin{pmatrix}
1&2&3&4&5&6&7\\
1&4&3&5&2&2&3
\end{pmatrix}.
\]
A bijective transformation is called a \e{permutation}.
Permutations can be written concisely using \e{disjoint cycle notation}; for example, $(1,2,4,5)(6,7)$ denotes the permutation
\[ \begin{pmatrix}
1&2&3&4&5&6&7\\
2&4&3&5&1&7&6
\end{pmatrix}. 
\]
Transformations can be \e{composed} using the $\circ$ operator; the image of $q$ under $s \circ t$ is $s(t(q))$.
A set of transformations of $Q$ that is closed under composition is called a \e{transformation monoid} on $Q$. 
The size of $Q$ is called the \e{degree} of the transformation monoid.
The \e{full transformation monoid} on $Q$ is the set of all transformations of $Q$. The \e{symmetric group} on $Q$ is the set of all permutations of $Q$.
A transformation monoid $M$ is \e{generated} by a set of transformations $T$ if every transformation in $M$ can be written as a composition of transformations from $T$.
We say a monoid is \e{$k$-generated} if it is generated by a set of size $k$.

Each DFAO $\cD = (Q,\Sig,\cdot,q_0,\Delta,\tau)$ has a transformation monoid associated with it, called the \e{transition monoid} of the DFAO. It is defined as follows. For each $w \in \Sig^*$, define the function $\ol{w} \co Q \ra Q$ by $\ol{w}(q) = q \cdot w$.
The function $\ol{w}$ is called the \e{action} of $w$ in $\cD$.
Composition of word actions obeys the following rule:
\[ \ol{x} \circ \ol{y} = \ol{yx},\quad\text{since $\ol{x}(\ol{y}(q)) = q \cdot y \cdot x = q \cdot yx = \ol{yx}$}. \]
Since the set $\{ \ol{w} : w \in \Sig^* \}$ of all word actions in $\cD$ is closed under composition, this set forms a transformation monoid on $Q$. This is the \e{transition monoid} of $\cD$.
The transition monoid is generated by the set $\{ \ol{a} : a \in \Sig\}$ of letter actions.

When working with multiple DFAOs, say $\cD = (Q,\Sig,\cdot,q_0,\Delta,\tau)$ and $\cD'=(Q',\Sig',\cdot',q'_0,\Delta',\tau')$, the notation $\ol{w}$ is ambiguous: it is unclear whether this is the action of $w$ in $\cD$ or in $\cD'$. We adopt the following convention: the notation $\ol{w}$ refers to the action of $w$ in a DFA whose transition function is named ``\,$\cdot$\,''. Thus in this case, $\ol{w}$ would refer to the action of $w$ in $\cD$, rather than $\cD'$. This convention will be sufficient to keep things unambiguous in this paper.

If $w = a_1a_2\dotsb a_{n-1} a_n$ is a word over $\Sig^*$ with $a_1,\dotsc,a_n \in \Sig$, the \e{reverse} of $w$ is the word $w^R = a_n a_{n-1} \dotsb a_2 a_1$. 
Observe that
\[ \ol{a_1} \circ \ol{a_2} \circ \dotsb \circ \ol{a_{n-1}} \circ \ol{a_n} = \ol{a_n a_{n-1} \dotsb a_2 a_1} = \ol{w^R}. \]
On the other hand,
\[ \ol{a_n} \circ \ol{a_{n-1}} \circ \dotsb \circ \ol{a_{2}} \circ \ol{a_1} = \ol{a_1a_2 \dotsb a_{n-1} a_n} = \ol{w}. \]
The \e{reverse} of a finite state function $f \co \Sig^* \ra \Delta$ is the function $f^R \co \Sig^* \ra \Delta$ defined by $f^R(w) = f(w^R)$. 
Following Allouche and Shallit~\cite[Theorem 4.3.3]{AlSh03}, we give a DFAO construction for $f^R$ in terms of a DFAO for $f$.
\bp
\label{prop:reversal}
Let $\cD = (Q,\Sig,\cdot,q_0,\Delta,\tau)$ be a DFAO computing the function $f$. There exists a DFAO $\cD^R$ computing $f^R$.
\ep
\bpf
Let $\cD^R = (\Delta^Q,\Sig,\odot,\tau,\Delta,\Omega)$, where:
\bi
\item
The state set is $\Delta^Q$, the set of all functions from $Q$ to $\Delta$.
\item
The initial state is $\tau \co Q \ra \Delta$, the output map of $\cD$.
\item
The transition function $\odot$ is defined as follows: $g \odot a = g \circ \ol{a}$, for $g \in \Delta^Q$ and $a \in \Sig$.
\item
The output map $\Omega \co \Delta^Q \ra \Delta$ is defined by $\Omega(g) = g(q_0)$.
\ei
By definition, the function computed by $\cD$ is $f(w) = \tau(q_0 \cdot w)$.
The function computed by $\cD^R$ is $\Omega(\tau \odot w) = (\tau \odot w)(q_0)$; we must show this equals $f^R(w) = f(w^R)$.
If $w = a_1a_2 \dotsb a_n$, then we have
\[
\tau \odot w 
= \tau \odot a_1 \odot a_2 \odot \dotsb \odot a_n
= \tau \circ \ol{a_1} \circ \ol{a_2} \circ \dotsb \circ \ol{a_n}
= \tau \circ \ol{w^R}. \]
It follows that 
\[ (\tau \circ \ol{w^R})(q_0) = \tau(\ol{w^R}(q_0)) = \tau(q_0 \cdot w^R) = f(w^R) = f^R(w) \]
as required. 
\epf
The \e{state complexity} of a finite state function is the size of a minimal DFAO computing the function.
If a function $f$ is computed by an $n$-state minimal DFAO (i.e., the function has state complexity $n$), Proposition \ref{prop:reversal} shows that the state complexity of $f^R$ is bounded above by $|\Delta|^n$, since the size of the state set $\Delta^Q$ of $\cD^R$ is $|\Delta|^{|Q|} = |\Delta|^n$.

The following proposition makes it easier to compute the state complexity of $f^R$. The analogous result for DFAs is known (e.g., see~\cite[Proposition 3]{JiSe12}).
\bp
\label{prop:reversal-distinguishable}
If $\cD$ is trim, then all states of $\cD^R$ are pairwise distinguishable.
\ep
\bpf
Let $g$ and $h$ be distinct states of $\cD^R$. There exists $q \in Q$ such that $g(q) \ne h(q)$.
Since $\cD$ is trim, $q$ is reachable. Choose $w \in \Sig^*$ such that $q_0 \cdot w^R = q$.
Observe that $\Omega(g \odot w) = (g \circ \ol{w^R})(q_0) = g(q_0 \cdot w^R) = g(q)$, and similarly $\Omega(h \odot w) = h(q)$.
Since $\Omega(g \odot w) \ne \Omega(h \odot w)$, $g$ and $h$ are distinguishable.
\epf
If we take $\cD^R$ and remove all unreachable states from it (which does not change the function computed), we obtain a DFAO for $f^R$ with all states reachable and every pair of distinct states distinguishable. By Proposition \ref{prop:min}, this is a minimal DFAO for $f^R$. Hence given a function $f$ computed by a trim DFAO $\cD$, to determine the state complexity of $f^R$, we can simply count the number of reachable states in $\cD^R$.

\section{Main Results}
We first prove an important proposition, which shows that the state complexity of reversal of DFAOs is completely determined by the transition monoid and the output map.
\bp
\label{prop:reversal-sc}
Let $\cD = (Q,\Sig,\cdot,q_0,\Delta,\tau)$ be a trim DFAO computing function $f$.
Let $M$ be the transition monoid of $\cD$.
The state complexity of $f^R$ is $|\tau M|$, where $\tau M = \{ \tau \circ \ol{w} : w \in \Sig^*\}$. 
\ep
\bpf
The DFAO $\cD^R = (\Delta^Q,\Sig,\odot,\tau,\Delta,\Omega)$ computes $f^R$.
By Proposition \ref{prop:reversal-distinguishable}, all states of $\cD^R$ are distinguishable, so the state complexity of $f^R$ is the number of reachable states in $\cD^R$.

Recall from the proof of Proposition \ref{prop:reversal} that $g \odot w = g \circ \ol{w^R}$ for $g \co Q \ra \Delta$ and $w \in \Sig^*$. In particular, since $\tau$ is the initial state of $\cD^R$, every reachable state of $\cD^R$ has the form $\tau \odot w = \tau \circ \ol{w^R}$.
Hence the set of reachable states of $\cD^R$ is $\{ \tau \circ \ol{w^R} : w \in \Sig^* \}$. 
But this is the same set as $\tau M = \{ \tau \circ \ol{w} : w \in \Sig^*\}$.
It follows that the number of reachable states in $\cD^R$ is precisely $|\tau M|$.
\epf
Recall that for $|\Delta| = 2$, DFAOs are essentially the same as DFAs (if we view the output map as a Boolean function telling us whether a state is final or non-final). Hence we have the following corollary:
\bc
Let $\cD = (Q,\Sig,\cdot,q_0,F)$ be a trim DFA recognizing language $L$.
Let $M$ be the transition monoid of $\cD$.
The state complexity of $L^R$ is $|\chi_F M|$, where $\chi_F \co Q \ra \{0,1\}$ is the characteristic function of $F$.
\ec
Despite the simple proof, we found this result rather surprising. We have not seen a similar characterization of the state complexity of DFA reversal anywhere in the literature.


Throughout the rest of this section, $Q$ and $\Delta$ will be finite sets with $|Q| = n$ and $|\Delta| = k$, the monoid $M$ will be a transformation monoid on $Q$, and $\tau \co Q \ra \Delta$ will be a surjective function. Note that the surjectivity of $\tau$ implies $|\Delta| \le |Q|$. It is fine to make this assumption, since if $|\Delta| > |Q|$ there are more possible outputs than there are states, and so we can shrink $\Delta$ without loss of generality.

\bt
\label{thm:reversal-sc}
Let $M$ be the full transformation monoid on $Q$. Then $|\tau M| = k^n$ for all surjective functions $\tau \co Q \ra \Delta$.
\bpf
It suffices to show that every function $h \co Q \ra \Delta$ lies in $\tau M$, i.e., every such function $h$ can be written as $\tau \circ g$ for some $g \co Q \ra Q$

For $q \in Q$, we define $g(q)$ as follows. Since $\tau$ is surjective, there exists $p_q \in Q$ such that $\tau(p_q) = h(q)$. Define $g(q) = p_q$. Then $(\tau \circ g)(q) = \tau(g(q)) = \tau(p_q) = h(q)$ for all $q \in Q$, so $\tau \circ g = h$ as required.
\epf
\et
\bc
\label{cor:reversal-sc}
Let $f$ be a finite state function computed by a minimal DFAO $\cD = (Q,\Sig,\cdot,q_0,\Delta,\tau)$ with $|\Delta| \le |Q|$ (i.e., $k \le n$). The state complexity of $f^R$ is at most $|\Delta|^{|Q|} = k^n$, and this bound can be reached when $|\Sig| \ge 3$.
\ec
\bpf
The upper bound on $f^R$ follows from the construction for $\cD^R$. It suffices to prove this bound can be reached.

It is well-known that the full transformation monoid on $Q$ can be generated by three elements: two generators of the symmetric group on $Q$, and a transformation of rank $|Q|-1$. If $Q = \{1,\dotsc,n\}$, an explicit example of three generators is
\[ f_1 = (1,2,\dotsc,n),\quad f_2 = (1,2), \quad f_3 = (1 \ra 2). \]
Here $(1 \ra 2)$ denotes the function that maps $1$ to $2$ and fixes all other elements.
Choose $\{a,b,c\} \subseteq \Sig$ and let $\cD$ be a DFAO such that $\ol{a} = f_1$, $\ol{b} = f_2$ and $\ol{c} = f_3$. Then the transition monoid $M$ of $\cD$ is the full transformation monoid. Furthermore, $\cD$ is trim (all states can be reached via $\ol{a}$). Hence Proposition \ref{prop:reversal-sc} applies. If we take the output map $\tau$ to be surjective, by Theorem \ref{thm:reversal-sc} we get that the state complexity of $f^R$ is $|\tau M| = k^n$, as required.
\epf

The rest of this section is devoted to the case where $|\Sig| = 2$, i.e., where the input alphabet of the DFAO is binary. This case is significantly more complicated and difficult than the $|\Sig| \ge 3$ case. Note that if $|\Delta| = 2$, this case is equivalent to studying reversal of ordinary DFAs with binary alphabets, and it is known for DFAs that the upper bound of $2^n$ is reachable~\cite{JiSe12}. Thus we will only be concerned with $|\Delta| \ge 3$.

Since the state complexity of DFAO reversal is completely determined by the transition monoid and output map, naturally there are connections between the $|\Sig| = 2$ case and the problem of finding the largest $2$-generated transformation monoids of a particular degree. This problem has been studied by Holzer and K\"onig~\cite{HoKo04} and Krawetz, Lawrence and Shallit~\cite{KLS05}.

Following Holzer and K\"onig, we define two families of monoids. First and most important are the $U_{\ell,m}$ monoids~\cite[Definition 5]{HoKo04}. The monoid $U_{\ell,m}$ is a transformation monoid on $Q = \{1,\dotsc,\ell+m\}$ defined as follows. Let $\alpha \co Q \ra Q$ be the permutation $(1,\dotsc,\ell)(\ell+1,\dotsc,\ell+m)$. 
A function $\gamma \co Q \ra Q$ belongs to $U_{\ell,m}$ if and only if it satisfies one of the following conditions:
\be
\item
There exists $i \ge 0$ such that $\gamma = \alpha^i$, that is, $\gamma = \alpha \circ \alpha \circ \dotsb \circ \alpha$ (where there are $i$ occurrences of $\alpha$).
\item
$\gamma(\{1,\dotsc,\ell\}) \cap \gamma(\{\ell+1,\dotsc,\ell+m\}) \ne \emp$, and there exists an element $i \in \{\ell+1,\dotsc,\ell+m\}$ such that $i$ is not in the image of $\gamma$.
\ee
If $1 < \ell < m$ and $\gcd(\ell,m) =1$, then $U_{\ell,m}$ can be generated by two elements~\cite[Theorem 8]{HoKo04}. Krawetz~\cite{Kra03} gives an explicit generating set: one of the generators is $\alpha$, and the other is $\beta \co Q \ra Q$, where
\[ \beta = \begin{pmatrix}
1&2&3&4&\dotsb&\ell+m-1&\ell+m\\
\ell+1&2&3&4&\dotsb&\ell+m-1&1
\end{pmatrix} \]
if $k = 2$ or $\ell$ is even, and otherwise
\[ \beta = \begin{pmatrix}
1&2&3&4&\dotsb&\ell+m-1&\ell+m\\
\ell+1&3&2&4&\dotsb&\ell+m-1&1
\end{pmatrix}. \]
Let $n = \ell+m$. For $n \ge 7$ and $n$ prime, Holzer and K\"onig proved that there exist $\ell$ and $m$ with $1 < \ell < m$ and $\gcd(\ell,m) = 1$ such that $U_{\ell,m}$ is the largest 2-generated transformation monoid~\cite[Theorem 15]{HoKo04}. They conjecture that this also holds when $n \ge 7$ and $n$ is not prime.

When $n \le 6$, the largest 2-generated transformation monoids belong to a different family: the $V^d_n$ monoids~\cite[Definition 16]{HoKo04}. Let $\alpha$ be the permutation $(1,2,\dotsc,n)$. 
A function $\gamma \co Q \ra Q$ belongs to $V^d_n$ if and only if it satisfies one of the following conditions:
\be
\item
There exists $i \ge 0$ such that $\gamma = \alpha^i$.
\item
There exist $i,j \in \{1,\dotsc,n\}$ such that $\gamma(i) = \gamma(j)$ and $j \equiv i+d \pmod{n}$.
\ee
For $2 \le n \le 6$, Holzer and K\"onig determined explicit generating sets for the largest 2-generated transformation monoids on $Q = \{1,\dotsc,n\}$, which are all $V^d_n$ monoids for some $d$. One of the generators is always $\alpha_n = (1,2,\dotsc,n)$. For $2 \le n \le 6$, the other generator $\beta_n$ is:
\[ 
\beta_2 = 
\begin{pmatrix}
1&2\\
1&1
\end{pmatrix},\quad
\beta_3 = 
\begin{pmatrix}
1&2&3\\
1&1&3
\end{pmatrix},\quad
\beta_4 = 
\begin{pmatrix}
1&2&3&4\\
1&1&4&3
\end{pmatrix}, \]
\[ 
\beta_5 = 
\begin{pmatrix}
1&2&3&4&5\\
1&1&4&5&3
\end{pmatrix},\quad
\beta_6 = 
\begin{pmatrix}
1&2&3&4&5&6\\
1&4&1&5&6&2
\end{pmatrix}. \]
Holzer and K\"onig also give a more general construction for 2-element generating sets of $V^d_n$ monoids~\cite[Theorem 18]{HoKo04}.

We did not try to prove this, but it seems that that $V^1_n$ monoids give rise to examples of DFAs (equivalently, DFAOs with $|\Delta| = 2$) which are witnesses for the maximal state complexity of DFA reversal. We verified computationally for $2 \le n \le 11$ that if $|\Delta| = 2$, then there exists a function $\tau \co Q \ra \Delta$ such that $|\tau(V^1_n)| = 2^n$. For odd $n$, it seems all surjective functions $\tau$ work, while for even $n$, there are only two functions that work (the rest give $|\tau(V^1_n)| = 2^n-2$). If $\Delta = \{1,2\}$, then the two functions that work for even $n$ are (in list notation) $[1,2,1,2,1,2,\dotsc,1,2]$ and $[2,1,2,1,2,1,\dotsc,2,1]$. In terms of DFAs, this corresponds to taking $\{1,3,5,7,\dotsc,n-1\}$ or $\{2,4,6,8,\dotsc,n\}$ as the final state set (assuming $\{1,\dotsc,n\}$ is the state set).
As mentioned, all of the above is conjectural and we have not attempted a proof.

With these definitions done, we return to the problem of computing worst-case state complexity of reversal for binary input alphabets.
First we consider the special case $|Q| = |\Delta|$. Here it turns out that the state complexity problem almost completely reduces to the 2-generated monoid problem:

\bt
\label{thm:kn}
Let $f$ be a finite state function computed by a minimal DFAO $\cD = (Q,\Sig,\cdot,q_0,\Delta,\tau)$ with $|\Sig| = 2$ and $|Q| = |\Delta| = n$.
Let $m_2(n)$ denote the size of the largest 2-generated transformation monoid on $Q = \{1,2,\dotsc,n\}$ that occurs as the transition monoid of some trim DFA.
The state complexity of $f^R$ is at most $m_2(n)$, and this bound is reachable.
\et
\bpf
Let $\Sig = \{a,b\}$. By assumption, we can construct a trim DFAO $\cD$ so that $\ol{a}$ and $\ol{b}$ generate a monoid of size $m_2(n)$. and let $\tau \co Q \ra \Delta$ be a bijection. 
By Proposition \ref{prop:reversal-sc}, the state complexity of $f^R$ is $|\tau M|$. But $\tau$ is a bijection, so $|\tau M| = |M| = m_2(n)$.
\epf
It may be the case that for some values of $n$, the largest transformation monoid on $\{1,2,\dotsc,n\}$ generated by two elements does \e{not} occur as the transition monoid of a trim DFA. Thus we do not quite get a complete reduction to the 2-generated monoid problem. It seems very unlikely that a monoid which is not a transition monoid of a trim DFA could be maximal, since this would mean there is some state $q$ for which the monoid contains no functions mapping a reachable state to $q$, which would exclude very many functions. Note that the $U_{\ell,m}$ and $V^d_n$ monoids do occur as transition monoids of trim DFAs.

It is well known that if $|Q| \ge 3$, the full transformation monoid on a finite set $Q$ cannot be generated by two elements. Hence $m_2(n)$ never reaches the upper bound of $|\Delta|^{|Q|} = n^n$ except when $|Q| = n =  2$.

Table \ref{tb:kn} shows the known values for $m_2(n)$ for $2 \le n \le 7$, taken from~\cite[Table 1]{HoKo04}. The value is not known for $n > 7$ except when $n$ is prime, in which case $m_2(n)$ is the size of the largest 2-generated $U_{\ell,m}$ monoid. The values of $n^n$ are also shown for comparison.
\begin{table}[h]
\[
\begin{array}{|c|c|c|c|c|c|c|}
\hline
n&2&3&4&5&6&7\\
\hline
m_2(n)&4&24&176&2110&32262&610871\\
\hline
n^n&4&27&256&3125&46656&823543\\
\hline
\end{array}
\]
\caption{Values of $m_2(n)$ for $2 \le n \le 7$.}
\label{tb:kn}
\end{table}

We now turn to the case where $|\Delta| < |Q|$. Our main result in this case is a formula for the size of $|\tau U_{\ell,m}|$, which in turn leads to a lower bound on the worst-case state complexity of $f^R$.

\bt
\label{thm:uml}
Let $|\Delta| = k$ and let $|Q| = \ell+m = n$, with $2 \le k < n$ and $1 \le \ell \le m$.
Define
\[ F(k,\ell,m) = 
\sum_{i=1}^{\ell} \binom{k}{i}i!\stir{\ell}{i}(k-i)^m. \]
\[ G(k,\ell,m) = \begin{cases}
\lcm(\ell,m), & \text{ if $k \ge 4$;}\\
m, & \text{ if $k = 3$;}\\
1, & \text{ if $k = 2$.}
\end{cases}
\]
There exists a function $\tau \co Q \ra \Delta$ such that
\[ |\tau U_{\ell,m}| =
k^n - F(k,\ell,m) + G(k,\ell,m).
\]
\et
The notation $\stir{\ell}{i}$ means the number of partitions of the set $\{1,\dotsc,\ell\}$ into $i$ parts (that is, a Stirling number of the second kind).
\bpf
We start with a brief outline of the proof strategy.
Without loss of generality, assume $\Delta = \{1,\dotsc,k\}$ and $Q = \{1,\dotsc,n=\ell+m\}$.
Define $F_{\ell,m} = \{ f \co Q \ra \Delta : f(\{1,\dotsc,\ell\}) \cap f(\{\ell+1,\dotsc,\ell+m\}) = \emp\}$.
\bi
\item
First, we show that $\Delta^Q = \tau U_{\ell,m} \cup F_{\ell,m}$ for certain $\tau$.
\item
After proving this, the inclusion-exclusion principle gives the formula
\[ k^n = |\Delta^Q| = |\tau U_{\ell,m}| + |F_{\ell,m}| - |\tau U_{\ell,m} \cap F_{\ell,m}|. \]
\item
We show that $|F_{\ell,m}| = F(k,\ell,m)$.
\item
We show that $|\tau U_{\ell,m} \cap F_{\ell,m}| = G(k,\ell,m)$.
\item
Rearranging the inclusion-exclusion formula above gives the result.
\ei
Let us show that for an appropriate choice of $\tau \co Q \ra \Delta$, we have $\Delta^Q = \tau U_{\ell,m} \cup F_{\ell,m}$. That is, every function from $Q$ to $\Delta$ lies in one of $\tau U_{\ell,m}$ or $F_{\ell,m}$.

Let $\alpha \co Q \ra Q$ be the permutation $(1,\dotsc,\ell)(\ell+1,\dotsc,\ell+m)$. 
We select $\tau$ with the following properties:
\bi
\item
$\tau \co Q \ra \Delta$ is surjective.
\item
$\tau(\{1,\dotsc,\ell\}) \cap \tau(\{\ell+1,\dotsc,\ell+m\}) = \emp$, that is, $\tau \in F_{\ell,m}$.
\item
There exist distinct $p,p' \in \{\ell+1,\dotsc,\ell+m\}$ such that $\tau(p) = \tau(p')$.
\item
The size of the set $\{\tau \circ \alpha^i : i \ge 0\}$ is precisely $G(k,\ell,m)$.
\ei
We demonstrate that such a function exists after this proof, in Lemma \ref{lem:func}. In that lemma, we will see existence of such a function requires $k < n$ and $\ell \le m$; this is the only place we use these hypotheses.

Now, let $g \co Q \ra \Delta$ be arbitrary. We will show that if $g$ is \e{not} in $F_{\ell,m}$, then it must be in $\tau U_{\ell,m}$, thus proving that $\Delta^Q = \tau U_{\ell,m} \cup F_{\ell,m}$. To show that $g \in \tau U_{\ell,m}$, we define a function $f \co Q \ra Q$ such that $f \in U_{\ell,m}$ and $\tau \circ f = g$.

Since $g \not\in F_{\ell,m}$, there exist distinct elements $r \in \{1,\dotsc,\ell\}$ and $r' \in \{\ell+1,\dotsc,\ell+m\}$ such that $g(r) = g(r')$. 
Since $\tau$ is surjective, there exists $s$ such that $\tau(s) = g(r)$.
Furthermore, we can choose $s$ so that $s \ne p'$.  
Indeed, if $p'$ is one of the possible choices for $s$, then by the fact that $\tau(p) = \tau(p')$, we can choose $s = p$ instead. 
Now, we define $f \co Q \ra Q$ for each $q \in Q$ as follows:
\bi
\item
If $q \in \{r,r'\}$, define $f(q) = s$.
\item
If $g(q) = \tau(p)$ and $q \not\in \{r,r'\}$, define $f(q) = p$.
\item
Otherwise, choose an element $q'$ such that $\tau(q') = g(q)$ (by surjectivity) and define $f(q) = q'$.
\ei
We verify in each case that $\tau \circ f = g$:
\bi
\item
If $q = r$, then $f(r) = s$, so $\tau(f(r)) = \tau(s) = g(r)$.
\item
If $q = r'$, then $f(q) = s$, and since $g(r) = g(r')$ we have $\tau(f(r')) = \tau(s) = g(r) = g(r')$.
\item
If $q \not\in \{r,r'\}$ and $g(q) = \tau(p)$, then $f(q) = p$, so $\tau(f(q)) = \tau(p) = g(q)$.
\item
Otherwise, we have $f(q) = q'$ such that $\tau(f(q)) = \tau(q') = g(q)$.
\ei
Now, we show that $f \in U_{\ell,m}$. First, note that there exist elements $r \in \{1,\dotsc,\ell\}$ and $r' \in \{\ell+1,\dotsc,\ell+m\}$ such that $f(r) = f(r')$. Next, observe that the element $p' \in \{\ell+1,\dotsc,\ell+m\}$ is not in the image of $f$. To see this, note that if we have $f(q) = p'$, then we have $\tau(f(q)) = \tau(p') = \tau(p)$. But $\tau(f(q)) = g(q)$, so this implies $g(q) = \tau(p)$. In the case where $g(q) = \tau(p)$, we defined $f(q) = p \ne p'$, so this is a contradiction. It follows that $f$ meets the conditions to belong to $U_{\ell,m}$.

This proves that if $g \co Q \ra \Delta$ is not in $F_{\ell,m}$, then $g \in \tau U_{\ell,m}$ and thus $\Delta^Q = \tau U_{\ell,m} \cup F_{\ell,m}$.
Next, we show that $|F_{\ell,m}| = F(k,\ell,m)$. 

Write $f \in F_{\ell,m}$ in list notation as $[a_1,a_2,\dotsc,a_\ell,b_1,b_2,\dotsc,b_m]$, where $f(i) = a_i$ and $f(\ell+i) = b_i$. For this function to lie in $F_{\ell,m}$, we must have the property that $\{a_1,a_2,\dotsc,a_\ell\} \cap \{b_1,b_2,\dotsc,b_m\} = \emp$. Note that since $F_{\ell,m}$ is a set of functions from $Q$ to $\Delta$, we have $\{a_1,\dotsc,a_\ell\},\{b_1,\dotsc,b_m\} \subseteq \Delta$. We count the number of distinct ``function lists'' in $F_{\ell,m}$  as follows:
\bi
\item
Fix a set $S \subseteq \Delta$ and assume $\{a_1,\dotsc,a_\ell\} = S$. Let $|S| = i$.
\item
In the first segment $[a_1,\dotsc,a_\ell]$ of the list, each $a_i$ can be a arbitrary element of $S$. However, since $\{a_1,\dotsc,a_\ell\} = S$, each element of $S$ must appear at least once in the list. Thus the first segment $[a_1,\dotsc,a_\ell]$ of the list represents a \e{surjective} function from $\{1,\dotsc,\ell\}$ onto $S$. Since $|S| = i$, the number of such surjective functions is $i!\stir{\ell}{i}$, where $\stir{\ell}{i}$ denotes a Stirling number of the second kind (the number of partitions of $\{1,\dotsc,\ell\}$ into $i$ parts).
\item
In the second segment $[b_1,\dotsc,b_m]$ of the list, each $b_i$ must be an element of $\Delta \setminus S$, since we want $\{a_1,\dotsc,a_\ell\} \cap \{b_1,\dotsc,b_m\} = \emp$. Since $|S| = i$ and $|\Delta| = k$, there are $k-i$ elements to pick from in $\Delta \setminus S$, and we need to choose $m$ of them. Thus there are $(k-i)^m$ choices for the second segment of the list.
\item
In total, for a fixed set $S$ of size $i$, there are $i!\stir{\ell}{i}(k-i)^m$ distinct lists with $\{a_1,\dotsc,a_k\} = S$.
\item
Now, we take the sum over all possible choices for the set $S$. Since $S = \{a_1,\dotsc,a_\ell\}$ and $S$ is non-empty, we have $1 \le |S| \le \ell$. For each set size $i$, there are $\binom{k}{i}$ ways to choose $S \subseteq \Delta$ with $|S| = i$. Thus the total number of functions in $F_{\ell,m}$ is
\[ \sum_{i=1}^\ell \binom{k}{i}i!\stir{\ell}{i}(k-i)^m = F(k,\ell,m). \]
\ei

Next, we show that $|\tau U_{\ell,m} \cap F_{\ell,m}| = G(k,\ell,m)$. We claim that
\[ \tau U_{\ell,m} \cap F_{\ell,m} = \begin{cases}
\emp, & \text{if $\tau \not\in F_{\ell,m}$;}\\
\{\tau \circ \alpha^i : i \ge 0\}, & \text{if $\tau \in F_{\ell,m}$.}
\end{cases} 
\]
Then the size equality with $G(k,\ell,m)$ follows from the properties of $\tau$. 

To see the claim, suppose that $\tau \circ g \in F_{\ell,m}$ for some $g \in U_{\ell,m}$.
Since $g \in U_{\ell,m}$, either $g = \alpha^i$ for some $i$, or there exists $p \in \{1,\dotsc,\ell\}$ and $q \in \{\ell+1,\dotsc,\ell+m\}$ such that $g(p) = g(q)$.
In the latter case, $\tau(g(p)) = \tau(g(q))$, which contradicts the assumption that $\tau \circ g$ is in $F_{\ell,m}$.
Hence $g = \alpha^i$ for some $i \ge 0$, and so $\tau \circ g = \tau \circ \alpha^i$.
Now, note that 
$\tau(\alpha^i(\{1,\dotsc,\ell\})) = \tau(\{1,\dotsc,\ell\})$, 
and
$\tau(\alpha^i(\{\ell+1,\dotsc,\ell+m\})) = \tau(\{\ell+1,\dotsc,\ell+m\})$.
Thus $\tau \circ \alpha^i$ is in $F_{\ell,m}$ if and only if $\tau$ is in $F_{\ell,m}$, and the claim follows.

Finally, we can conclude the proof.
Recall that $|\Delta| = k$ and $|Q| = n$, and thus $|\Delta^Q| = |\Delta|^{|Q|} = k^n$.
Thus by the inclusion-exclusion principle, we have 
\[ k^n = |\Delta^Q| = |\tau U_{\ell,m}| + |F_{\ell,m}| - |\tau U_{\ell,m} \cap F_{\ell,m}|. \]
Rearranging this, we get:
\[ |\tau U_{\ell,m}| = k^n  - |F_{\ell,m}| + 
|\tau U_{\ell,m} \cap F_{\ell,m}|. \]
We proved that $|F_{\ell,m}| = F(k,\ell,m)$ and $|\tau U_{\ell,m} \cap F_{\ell,m}| = G(k,\ell,m)$. It follows that
$|\tau U_{\ell,m}| = k^n  - F(k,\ell,m) + G(k,\ell,m)$,
as required.
\epf

This theorem gives the following lower bound on the worst-case state complexity of DFAO reversal when $|\Sig| = 2$.
\bc
\label{cor:uml}
Let $|Q| = n \ge 2$ and $|\Delta| = k \ge 2$.
There exists a trim DFAO $\cD = (Q,\Sig,\cdot,q_0,\Delta,\tau)$ computing function $f$, with $|\Sig| = 2$ and $k < n$, such that the state complexity of $f^R$ is
\[ \max\{ k^n - F(k,\ell,m) + G(k,\ell,m) : 1 < \ell < m, \ell+m = n, \gcd(\ell,m)=1\}. \]
\ec
\bpf
Pick $\ell$ and $m$ such that $1 < \ell < m$, $\ell+m = n$ and $\gcd(\ell,m) = 1$.
Then $U_{\ell,m}$ can be generated by two elements. Hence we can construct a DFAO $\cD$ over a binary alphabet with state set $Q = \{1,\dotsc,n\}$ and transition monoid $U_{\ell,m}$. This DFAO will be trim: all states in $\{1,\dotsc,\ell\}$ are reachable by $\alpha = (1,\dotsc,\ell)(\ell+1,\dotsc,\ell+m)$, and $U_{\ell,m}$ contains elements which map $1$ to $\ell+1$, so the rest of the states are reachable. By Theorem \ref{thm:uml}, there exists $\tau \co Q \ra \Delta$ such that 
\[ |\tau U_{\ell,m}| = k^n - F(k,\ell,m) + G(k,\ell,m). \]
Take $\tau$ as the output map of $\cD$.
Then by Proposition \ref{prop:reversal-sc}, the state complexity of $f^R$ is $|\tau U_{\ell,m}|$. Taking the maximum over all values of $\ell$ and $m$ that satisfy the desired properties gives the result.
\epf
Table \ref{tb:uml} gives the values of this lower bound for various values of $|\Delta| = k$ and $|Q| = n$ with $k < n$. Note that for $n \in \{1,2,3,4,6\}$ there are no pairs $(\ell,m)$ such that $1 < \ell < m$, $\ell+m = n$ and $\gcd(\ell,m) = 1$, so those values of $n$ are ignored.
\begin{table}[h]
\[
\begin{array}{|c|c|c|c|c|c|}
\hline
k\backslash n 
 & 5  & 6 & 7   & 8   & 9   \\
\hline
2& 31 & - & 127 & 255 & 511 \\
\hline
3& 216& - & 2125& 6452&19550 \\
\hline
4& 826& - & 15472 & 63403 & 258360 \\
\hline
5& - & - & 71037 & 368020 & 1902365 \\
\hline
6& - & - & 243438 & 1539561 & 9657446 \\
\hline
\end{array}
\]
\caption{Values for the lower bound of Corollary \ref{cor:uml}.}
\label{tb:uml}
\end{table}
Note that for $|\Delta| = 2$, this lower bound is off by one from the upper bound of $2^n$. The known examples where the upper bound $2^n$ is achieved do not use $U_{\ell,m}$ monoids.

We suspect the lower bound of Corollary \ref{cor:uml} may be optimal for $n \ge 7$. We were unable to find any examples exceeding the bound through computational experiments (discussed in Section \ref{sec:comp}). 

The case of $n=5$, which is the only case below $7$ where our lower bound is defined, is rather interesting.
Holzer and K\"onig proved by brute force search that the largest 2-generated transformation monoid of degree 5 is $V^1_5$, so one might expect that the maximal values of $|\tau M|$ in the $n=5$ case would be given by taking $M = V^1_5$. Indeed, for $(k,n) = (3,5)$, the true maximum is $218$, and this is achieved by $|\tau V^1_5|$ with $\tau = [1,2,1,2,3]$. However, for $(k,n) = (4,5)$: the value $826$ is achieved by $|\tau U_{2,3}|$ with $\tau = [1,2,3,4,4]$, while the maximal value of $|\tau V^1_5 |$ over all $\tau$ and $d$ is $789$, despite the fact that $|U_{2,3}| = 1857$ and $|V^1_5| = 2110$. Thus maximal monoids $M$ do not necessarily give the maximal values for $|\tau M|$.

In the proof of Theorem \ref{thm:uml}, we used the fact that a function with certain properties exists. We now give a rather tedious proof of this fact.

\bl
\label{lem:func}
Let $\Delta = \{1,\dotsc,k\}$ and let $Q = \{1,\dotsc,n\}$, with $2 \le k < n$.
Fix $\ell$ and $m$ such that $\ell+m = n$ and $1 \le \ell \le m$. 
Let $\alpha \co Q \ra Q$ be the permutation $\alpha = (1,\dotsc,\ell)(\ell+1,\dotsc,\ell+m)$. 
There exists a function $\tau \co Q \ra \Delta$ with the following properties:
\bi
\item
$\tau \co Q \ra \Delta$ is surjective.
\item
$\tau(\{1,\dotsc,\ell\}) \cap \tau(\{\ell+1,\dotsc,\ell+m\}) = \emp$.
\item
There exist distinct $p,p' \in \{\ell+1,\dotsc,\ell+m\}$ such that $\tau(p) = \tau(p')$.
\item
The size of the set $\{\tau \circ \alpha^i : i \ge 0\}$ is precisely given by the function $G(k,\ell,m)$ defined in Theorem \ref{thm:uml}.
\ei
\el
\bpf
For $|\Delta| = k \ge 3$ and $(\ell,m) \ne (2,2)$, we define $\tau$ as follows:
\be
\item
We partition $\Delta$ into two sets $L$ and $M$, with $|L| = \min\{k-2,\ell\}$.   
\item
(a) If $|L| = \ell$, let $L = \{1,\dotsc,\ell\}$ and define $\tau(i) = i$ for $1 \le i  \le \ell$.\\
(b) If $|L| = k-2 < \ell$, write $L = \{1,\dotsc,k-2\}$ and define $\tau(i) = i$ for  $1 \le i \le k-2$. Define $\tau(i) = k-2$ for $k-1 \le i \le \ell$.
\item
Consider $|M| = k-|L|$.\\
(a) Suppose $k-2 \ge \ell$, and thus $|L| = \ell$.
Then $\ell+2 \le k < n = \ell+m$, so $2 \le k-\ell < m$.
Thus $2 \le |M| < m$.\\
(b) Suppose $k-2 < \ell$, and thus $|L| = k-2$.
Then $|M| = k-\ell = 2$.
\item
(a) If $|L| = \ell$, we have $M = \{\ell+1,\dotsc,\ell+j\}$, where $j < m$ is such that $\ell+j = k$.
Define $\tau(\ell+i) = \ell+i$ for $1 \le i \le j$. Define $\tau(\ell+i) = \ell+j = k$ for $j+1 \le i \le m$.\\
(b) If $|L| = k-2$, we have $M = \{k-1,k\}$. Define $\tau(\ell+1) = k-1$ and $\tau(\ell+i) = k$ for $2 \le i \le m$.
\ee
To illustrate this construction, we give three examples of $\tau$ for different values of $k$, $\ell$ and $m$.
If $k = 6$, $\ell = 3$ and $m = 5$, we partition $\Delta = \{1,2,3,4,5,6\}$ into $L = \{1,2,3\}$ and $M = \{4,5,6\}$, and we get
\[ \tau = \begin{pmatrix}
1 & 2 & 3 & 4 & 5 & 6 & 7 & 8 \\
1 & 2 & 3 & 4 & 5 & 6 & 6 & 6
\end{pmatrix}.
\]
If $k = 5$, $\ell = 4$ and $m = 5$, we partition $\Delta = \{1,2,3,4,5\}$ into $L = \{1,2,3\}$ and $M = \{4,5\}$, and we get
\[ \tau = \begin{pmatrix}
1 & 2 & 3 & 4 & 5 & 6 & 7 & 8 & 9 \\
1 & 2 & 3 & 3 & 4 & 5 & 5 & 5 & 5
\end{pmatrix}.
\]
If $k = 3$, $\ell = 4$ and $m = 4$, we partition $\Delta = \{1,2,3,4,5\}$ into $L = \{1\}$ and $M = \{2,3\}$, and we get
\[ \tau = \begin{pmatrix}
1 & 2 & 3 & 4 & 5 & 6 & 7 & 8 \\
1 & 1 & 1 & 1 & 2 & 3 & 3 & 3
\end{pmatrix}.
\]
Note that in all cases, $\tau(\{1,\dotsc,\ell\}) = L$ and $\tau(\{\ell+1,\dotsc,\ell+m\}) = M$.

This covers the definition of $\tau$ for $k \ge 3$ and $(\ell,m) \ne (2,2)$.
In the special case where $k \ge 3$ and $(\ell,m) = (2,2)$, the fact that $k < n = \ell+m = 4$ implies $k=3$.
Here we define
\[ \tau = \begin{pmatrix} 1 & 2 & 3 & 4 \\ 1 & 2 & 3 & 3 \end{pmatrix}. \]
Finally, for $k = 2$, we define $\tau$ by $\tau(i) = 1$ for $1 \le i \le \ell$ and $\tau(i) = 2$ for $\ell+1 \le i \le \ell+m$.

We now demonstrate $\tau$ has all the desired properties. The surjectivity of $\tau$ and the fact that $\tau(\{1,\dotsc,\ell\}) \cap \tau(\{\ell+1,\dotsc,\ell+m\}) = \emp$ can be easily verified by a close reading of the definition. 

Consider the third property: there exist distinct $p,p' \in \{\ell+1,\dotsc,\ell+m\}$ such that $\tau(p) = \tau(p')$. We can see this from the construction of $\tau$ as follows:
\bi
\item
For $k \ge 3$ and $(\ell,m) \ne (2,2)$, observe that we have $\tau(\{\ell+1,\dotsc,\ell+m\}) = M$.
If $|L| = \ell$, we have $|M| < m$, so $\tau$ must identify two elements of $\{\ell+1,\dotsc,\ell+m\}$.
If $|L| = k-2$, then $|M| = 2$, so we have $|M| < m$ in all cases except $m = 2$. 
But if $m = 2$, then $\ell \le m$ implies $\ell \le 2$.
Since $(\ell,m) \ne (2,2)$, we must have $\ell = 1$; but the fact that $k < n = \ell+m$ implies $k < 3$, which is a contradiction.
So if $k \ge 3$ and $(\ell,m) \ne (2,2)$, then $\{\ell+1,\dotsc,\ell+m\}$ gets mapped onto a set of size less than $m$ by $\tau$.
\item
For $k \ge 3$ and $(\ell,m) = (2,2)$, we see that $\tau(3) = \tau(4)$.
\item
For $k = 2$, by the fact that $k < n = \ell+m $ we must have $m \ge 2$, and so $\tau(\ell+1) = \tau(\ell+m) = 2$ and $\ell+1 \ne \ell+m$.
\ei

Finally, consider the last property: the set $\{\tau \circ \alpha^i : i \ge 0\}$ has size $G(k,\ell,m)$. That is, it has size $\lcm(\ell,m)$ if $k \ge 4$, size $m$ if $k = 3$, and size 1 if $k = 2$. 

Write $\tau$ in list notation as $[a_1,a_2,\dotsc,a_n]$, where $\tau(i) = a_i$ for $1 \le i \le n$. Note that $L = \{a_1,\dotsc,a_\ell\}$ and $M = \{a_{\ell+1},\dotsc,a_{\ell+m}\}$.
Observe that the list notation for $\tau \circ \alpha$ is 
$[a_{\alpha(1)},a_{\alpha(2)},\dotsc,a_{\alpha(n)}]$.
Similarly, the list notation for $\tau \circ \alpha^i$ is 
$[a_{\alpha^i(1)},a_{\alpha^i(2)},\dotsc,a_{\alpha^i(n)}]$.
It follows that the number of distinct ``function lists'' in the set $\{\tau \circ \alpha^i : i \ge 0\}$ is bounded by the order of the permutation $\alpha$, which is $\lcm(\ell,m)$.

For $k \ge 4$, we will show that all $\lcm(\ell,m)$ of these lists are distinct.
To see this, let $q$ be the smallest element of $M$, and consider where the values $1$ and $q$ appear in the list
$[a_{\alpha^i(1)},a_{\alpha^i(2)},\dotsc,a_{\alpha^i(n)}]$.
By the definition of $\alpha$,
the value $1$ must appear at some position $p_i$ with $1 \le p_i \le \ell$ and $p_i+i \equiv 1 \pmod{\ell}$.
Similarly, the value $q$ must appear some position $r_i$ with $\ell+1 \le r_i \le \ell+m$ and $r_i+i \equiv q \pmod{m}$. 
Notice that by the definition of $\tau$, the elements $1$ and $q$ have unique preimages under $\tau$, and thus each only appears once in the list.

We claim that if $i \not\equiv j \pmod{\lcm(\ell,m)}$, then $(p_i,r_i) \ne (p_j,r_j)$. 
To see this, suppose for a contradiction that $(p_i,r_i) = (p_j,r_j)$. 
Since $p_i = p_j$, and we have $p_i+i \equiv p_j+j \equiv 1 \pmod{\ell}$, it follows that $i \equiv j \pmod{\ell}$. Similarly, we have $i \equiv j \pmod{m}$. 
Write $i = j+x\ell$ for some integer $x$; then we have $j+x\ell \equiv j \pmod{m}$.
It follows $x\ell \equiv 0 \pmod{m}$ and thus $m$ divides $x\ell$. Since $m$ and $\ell$ both divide $x\ell$, it follows that $\lcm(\ell,m)$ divides $x\ell$, and so we can write $x\ell = y\lcm(m,\ell)$ for some integer $y$.
Thus $i = j+y\lcm(m,\ell)$ and it follows that $i \equiv j \pmod{\lcm(m,\ell)}$.

This proves that if $\alpha^i \ne \alpha^j$, then the positions of $1$ and $q$ in the lists for $\tau \circ \alpha^i$ and $\tau \circ \alpha^j$ will be different. Thus $\tau \circ \alpha^i \ne \tau \circ \alpha^j$, and it follows the size of $\{\tau \circ \alpha^i : i \ge 0\}$ is precisely the order of the permutation $\alpha$, which is $\lcm(\ell,m)$.

This deals with the $k \ge 4$ case. For $k = 3$, first suppose $(\ell,m) \ne (2,2)$.
Since $k = 3$, we have $k-2 = 1$, so we always have $\min\{k-2,\ell\} = k-2$. In this case, recall that we have $\tau(i) = k-2 = 1$ for $1 \le i \le \ell$.

It follows that $L = \{a_1,\dotsc,a_\ell\} = \{1\}$, so the list notation for $\tau \circ \alpha^i$ is $[1,1,\dotsc,1,a_{\alpha^i(\ell+1)},\dotsc,a_{\alpha^i(\ell+m)}]$. Hence in this case, we get $m$ distinct lists, corresponding to the $m$ possible positions of $q$ (which still has a unique preimage under $\tau$) in the second part of the list. 

For $k = 3$ and $(\ell,m) = (2,2)$, it is easy to see we get $m=2$ lists: $[1,2,3,3]$ and $[2,1,3,3]$.
Finally, for $k = 2$, we just have one list $[1,1,\dotsc,1,2,2,\dotsc,2]$, where there are $\ell$ 1's and $m$ 2's.
This proves that $\{\tau \circ \alpha^i : i \ge 0\} = G(k,\ell,m)$ in all cases, and thus completes the proof of the lemma. 
\epf

\section{Computational Experiments}
\label{sec:comp}
We performed two types of computational experiments to help determine the worst-case state complexity of DFAO reversal: brute force searches and random searches. The goal of these experiments was to find, for various values of $|Q| = n $ and $|\Delta| = k$, the maximal size of $|\tau M|$, where $M$ is a monoid generated by two functions $\alpha \co Q \ra Q$ and $\beta \co Q \ra Q$, and $\tau \co Q \ra \Delta$ is a surjective function.

For our ``brute force'' searches, we did not actually test all triples $(\alpha,\beta,\tau)$. Several observations allowed us to reduce the search space. 

First, we need only test monoids up to conjugate isomorphism.
If monoids $M$ and $N$ are conjugate, we claim there exist functions $\tau$ and $\tau'$ such that $|\tau M| = |\tau' N|$.
Indeed, write $\tau M = \{ \tau \circ \rho : \rho \in M\}$
If $N = \gamma^{-1}M\gamma$ for a permutation $\gamma$, then $\tau' N = \{ \tau' \circ \gamma^{-1} \circ \rho \circ \gamma : \rho \in M\}$.
Thus taking $\tau' = \gamma^{-1} \circ \tau \circ \gamma$ gives
$(\gamma^{-1} \circ \tau \circ \gamma )N = \{ \gamma^{-1} \circ \tau \circ \rho \circ \gamma : \rho\in M\}$, which has the same size as $\tau M$. 
It follows that as long as we test every function $\tau$ for each monoid, we only need to test monoids up to conjugate isomorphism to find maximal values for $|\tau M|$.

Second, we may assume one of $\alpha$ or $\beta$ is a permutation.
Holzer and K\"onig show that if $M$ is generated by two non-permutations, then $M$ is conjugate-isomorphic to a submonoid of $V^1_{n}$~\cite[Lemma 25]{HoKo04}. Hence it suffices to test $V^1_{n}$ instead of $M$, and $V^1_{n}$ can be generated by a permutation and a non-permutation. Without loss of generality, we will assume $\alpha$ is a permutation.

Finally, we do not need to test every transformation $\beta$; we just need to test one transformation from each conjugacy class of the full transformation monoid. 
Indeed, let $\{C_1,\dotsc,C_m\}$ be the conjugacy classes and let $\beta_i$ be a representative of $C_i$. 
If $\beta$ lies in class $C_i$, there exists a permutation $\gamma$ such that $\gamma \circ \beta \circ \gamma^{-1} = \beta_i$. 
Hence the monoid $M$ generated by $\alpha$ and $\beta$ is conjugate-isomorphic to the monoid $N$ generated by $\gamma \circ \alpha \circ \gamma^{-1}$ and $\beta_i$, so it suffices to just test $N$.
This is a significant reduction; for example, there are $6^6 = 46656$ transformations on a set of size six, but only 130 distinct conjugacy classes. The lists of conjugacy class representatives we used were obtained from Mitchell~\cite{JDM16}.

To summarize, we tested all triples $(\alpha,\beta,\tau)$ where:
\bi
\item
$\alpha$ ranges over all permutations of $Q$;
\item
$\beta$ ranges over a full set of conjugacy class representatives for the full transformation monoid on $Q$;
\item
$\tau$ ranges over all surjective functions $\tau \co Q \ra \Delta$.
\ei

Even with these reductions, the exponential nature of this problem means that brute force searches become unreasonably slow very quickly. Thus for larger values of $|\Delta|$ and $|Q|$, we performed random searches. For these searches, we used the same search space as for the brute force searches but simply picked elements at random repeatedly and kept track of the largest value of $|\tau M|$ discovered. This allowed us to obtain some evidence for our conjectures even in the cases where a complete brute force search was infeasible.

The results of our experiments are shown in Table \ref{tb:comp}. The values in {\bf bold} are true maximal values for $|\tau M|$ (and thus for the state complexity of binary DFAO reversal), which have been confirmed by brute force search. The other values in the table are simply the largest we found through random search, though we conjecture these are maximal as well. 

\begin{table}[h]
\[
\begin{array}{|c|c|c|c|c|c|c|}
\hline
k \backslash n&
 3&4&5&6&7&8\\
\hline
3&\mathbf{24}&\mathbf{67}&\mathbf{218}&\mathbf{699}&2125&6452\\
\hline
4&-&\mathbf{176}&\mathbf{826}&3526&15472&63403\\
\hline
\end{array}
\]
\caption{Largest known values for $|\tau M|$, where $M$ is a 2-generated transformation monoid on $\{1,\dotsc,n\}$ and $\tau \co \{1,\dotsc,n\} \ra \{1,\dotsc,k\}$ is surjective.}
\label{tb:comp}
\end{table}

Note that for $n \ge 7$, the conjectured maximal values in Table \ref{tb:comp} match the values in Table \ref{tb:uml} for lower bound of Corollary \ref{cor:uml}. For this reason, we suspect the bound of Corollary \ref{cor:uml} may in fact be optimal for $n \ge 7$. 

\section{Conclusions}
For DFAs, the worst-case state complexity of the reversal operation is $2^n$ for languages of state complexity $n$. When we generalize to DFAOs, the worst-case state complexity is bounded above by $k^n$, where $k$ is the number of outputs of the DFAO. We proved that this upper bound can be attained by DFAOs over a ternary alphabet. For binary alphabets, we demonstrated there are connections with the problem of finding the largest 2-generated transformation monoid, and gave a lower bound on the worst-case state complexity for the $k < n$ case.

We state some open problems arising from this work.
\be
\item
Is the upper bound of $k^n$ is reachable over a binary alphabet when $k \ge 3$? We strongly suspect it is not. This is proven for $k = n$, and for $k < n$ we have verified computationally that it is not reachable when $(k,n) \in \{(3,4),(3,5),(3,6),(4,5)\}$.
\item
Is the lower bound given in Corollary \ref{cor:uml} optimal for $n \ge 7$? We suspect that it is, but we currently have limited evidence, so we would not be hugely surprised if a counterexample was found.
\item
Does the largest 2-generated transformation monoid always occur as the transition monoid of some trim DFA? This would mean when $k = n$, the problem of finding the worst-case state complexity of DFAO reversal reduces to the problem of finding the largest 2-generated transformation monoid. (See Theorem \ref{thm:kn} and the discussion afterwards.)
\item
For the brute force searches described in Section \ref{sec:comp}, can we make further reductions to the search space, allowing more values to be computed? 
\item
For reversal of ordinary DFAs, the ``magic number'' problem has been studied. A natural number $\alpha$ with $\log_2 n \le \alpha \le 2^n$ is called \e{magic} (for the DFA reversal operation) if there does not exist a language $L$ of state complexity $n$ such that $L^R$ has state complexity $\alpha$. The range $\log_2 n \le \alpha \le 2^n$ is chosen since these are the lower and upper bounds for state complexity of DFA reversal. It is known that there are no magic numbers for reversal of DFAs~\cite{Jir14}. Do magic numbers exist for reversal of DFAOs?
\ee

\subsection*{Acknowledgements}
I thank Janusz Brzozowski and Jeffrey Shallit for proofreading and helpful comments.  
This work was supported by the Natural Sciences and Engineering
Research Council of Canada under grant No.\ OGP0000871.


%

\bibliographystyle{abbrv}
\bibliography{reversal-new}

\end{document}